\documentclass[%
 aip,
 amsmath,amssymb,
 reprint,%
]{revtex4-1}

\usepackage{graphicx}% Include figure files
\usepackage{dcolumn}% Align table columns on decimal point
\usepackage{bm}% bold math
\usepackage[utf8]{inputenc}
\usepackage[T1]{fontenc}
\usepackage{mathptmx}
\usepackage{etoolbox}
\usepackage{cancel}
\usepackage[normalem]{ulem}
\usepackage[separate-uncertainty = true,multi-part-units=single]{siunitx}
\DeclareSIUnit\bar{bar}
\newcommand{\dd}{\mathrm{d}}
\DeclareSIPower\quartic\tothefourth{4}
\DeclareSIUnit\angstrom{\text{Å}}

\makeatletter
\def\@email#1#2{%
 \endgroup
 \patchcmd{\titleblock@produce}
  {\frontmatter@RRAPformat}
  {\frontmatter@RRAPformat{\produce@RRAP{*#1\href{mailto:#2}{#2}}}\frontmatter@RRAPformat}
  {}{}
}%
\makeatother
\begin{document}

%\preprint{AIP/123-QED}

\title[~]{Diffusiophoretic migration of colloidal particles in  sucrose gradients}
% Force line breaks with \\

\author{Antoine Monier}
\affiliation{CNRS, Syensqo, LOF, UMR 5258, Université de Bordeaux, 178 av. Schweitzer, 33600 Pessac, France}
\author{Brielle Byerley}
\affiliation{Department of Physics and Bioengineering, Lehigh University, Bethlehem, Pennsylvania, 18015, United States}
\author{Julien Renaudeau}
\affiliation{CNRS, Syensqo, LOF, UMR 5258, Université de Bordeaux, 178 av. Schweitzer, 33600 Pessac, France}
\author{H. Daniel Ou-Yang}
\affiliation{Department of Physics and Bioengineering, Lehigh University, Bethlehem, Pennsylvania, 18015, United States}
\author{Pierre Lidon}
\affiliation{CNRS, Syensqo, LOF, UMR 5258, Université de Bordeaux, 178 av. Schweitzer, 33600 Pessac, France}
\author{Jean-Baptiste Salmon}
\affiliation{CNRS, Syensqo, LOF, UMR 5258, Université de Bordeaux, 178 av. Schweitzer, 33600 Pessac, France}

%\author{Julien Renaudeau}
%\author{Pierre Lidon}%
%\author{Jean-Baptiste Salmon}
%\affiliation{CNRS, Syensqo, LOF, UMR 5258, Université de Bordeaux, 178 av. Schweitzer, 33600 Pessac, France.}%
 
\date{\today}% It is always \today, today,
             %  but any date may be explicitly specified

\begin{abstract}
Diffusiophoresis (DP) refers to the migration  of particles driven by a solute concentration gradient in a liquid. Observations in the case of molecular neutral solutes are rather scarce, due to the low drift velocities in dilute solutions, and the difficulty in distinguishing DP from other phenomena in concentrated solutions.      
We investigated experimentally DP of dispersed colloids driven by concentration gradients of sucrose in water at relatively high concentrations, $C \simeq \SI{1}{\mol\per\liter}$. More precisely, we designed a microfluidic chip to impose a time-dependent sucrose gradient in dead-end microchannels with minimized parasitic flows. Significant migration of the particles toward the regions of low sucrose concentration has been observed,  with  velocities up to a few $\SI{}{\micro\meter\per\second}$. Particle tracking and Raman confocal spectroscopy  were used to measure individual trajectories and the unsteady sucrose concentration profile respectively. The latter is correctly described by a diffusion equation, but with an interdiffusion coefficient that significantly depends on $C$ in the range of concentrations investigated. We then showed that a model of DP based on a steric exclusion of sucrose molecules from the particle surface with an exclusion length $R_i = \SI{5 \pm 0.9}{\angstrom}$ (close to the characteristic size of  the sucrose molecule),  accounts for the observed trajectories. Possible sources for the observed  scattering of our experimental data are finally discussed: Brownian motion and  advection of the particles by bulk flows driven by diffusioosmosis  at the channel walls and buoyancy.
\end{abstract}

\maketitle
%%%%%%%%%%%%%%%%%%%%%%%%%%%%%%%%%%%%%%%%%%%%%%%%%%%%%%%%%%%%%%%%%%%%%
%%%%%%%%%%%%%%%%%%%%%%%%%%%%%%%%%%%%%%%%%%%%%%%%%%%%%%%%%%%%%%%%%%%%%

\section{Introduction}\label{sec:intro}

Diffusiophoresis refers to the relative motion between a solid and a liquid induced by a solute concentration gradient, due to the surface-solute interaction~\cite{Derjaguin1993}.
For a fixed solid wall, this phenomenon can be seen as a {\it slip} velocity $v_s$ along the gradient and generates a volume flow, called diffusioosmosis (DO). For dispersed objects in a fluid, it leads to their migration, called diffusiophoresis (DP).
These interfacial transport phenomena, early revealed by Derjaguin {\it et al.}~\cite{Derjaguin1993}, then  studied theoretically by Anderson and collaborators~\cite{Anderson1982,Anderson1989} and later by Bocquet and collaborators~\cite{Marbach2017,Marbach2020}, have recently received a particular surge of interest with the advent of microfluidics~\cite{Velegol2016,Marbach2019,Shim2022,Ault2025}. 
For dispersed colloids in a confined geometry, with a solute concentration gradient driving both DO at the walls and  the colloid/fluid interfaces, the motion of colloids results from the combination of DP migration and advection by  bulk DO flow, which can be difficult to decipher from each other.
Most studies focused on gradients of electrolytes, leading to  DP migration of charge-stabilized colloids  and DO  flows in microchannels~\cite{Velegol2016,Marbach2019,Shim2022,Ault2025,Abecassis:08,Palacci2010,McDermott2012,Kar2015,Liu2025}.  This is mainly because the long-range  electrostatic interaction combined with an electrophoretic contribution  to DO/DP lead to  significant  velocities, up to a few $\SI{}{\um\per\second}$, with  moderate gradients in dilute solutions 
($<\SI{10}{\milli\mol\per\liter}$).
Since the interaction potential between electrolytes and charged surfaces is well known, these studies have made it possible to obtain quantitative descriptions and even to test predictions in regimes beyond the standard models~\cite{Shi2016,Gupta2019,Alessio2021,Shin2016,Rasmussen2020,NeryAzevedo:17,Akdeniz2023,Williams2024,Shah2022,Jotkar2024rev,Jotkar2024}.

In contrast, the case of neutral solutes has been significantly less studied, 
see, for instance, Refs.~\cite{Staffeld1989,Lee2014,Paustian:15,Lee2017,Williams2020,Nguyen2022,Akdeniz2024} for experimental studies. Theoretically~\cite{Anderson1982,Marbach2017}, in the dilute regime, the DO slip velocity $v_s$ is expected to be proportional to the gradient of solute concentration $\nabla C$ (expressed in $\SI{}{\mole\per\quartic\meter}$):
\begin{eqnarray}
    v_s = - \Gamma R T \nabla C, 
    \label{eq:DP_general_conc}
\end{eqnarray}
\noindent in which $R$ is the ideal gas constant, $T$ is the absolute temperature and $\Gamma$ (expressed in $\SI{}{\meter\squared\per\pascal\per\second}$) is the so-called diffusioosmotic mobility, that depends on the details of the solute/surface interaction. 
The slip is directed toward high (respectively low) solute concentration in the case of repulsive (respectively attractive) potential, for which $\Gamma <0$ (respectively $\Gamma >0$). 
For suspended colloids of radius $a$, the DP drift velocity $v_\mathrm{DP}$ is given by $-v_s$ provided that $a$ is larger than the range of the interaction potential~\cite{Anderson1982}, and particles migrate 
towards low  (resp. high) solute concentration in the case of repulsive (resp. attractive) interaction.
Note that for strongly attractive potentials, deviations from eq~\eqref{eq:DP_general_conc} are expected because of polarization of the solute diffuse layer at the colloid surface and/or because convection for the solute transport cannot be neglected~\cite{Anderson1991,Keh2001}.

For solutes that are sterically excluded from a solid surface, the aforementioned deviations are not present and the diffusioosmotic mobility is theoretically given in 
dilute solution by~\cite{Anderson1982,Marbach2017}:
\begin{eqnarray}
    \Gamma = - \frac{R_i^2}{2\eta_w}, 
    \label{eq:Gamma_HS}
\end{eqnarray}
for a hard-sphere potential of range $R_i$, with $\eta_w$ the viscosity of water.
 Eq~\eqref{eq:DP_general_conc} and eq~\eqref{eq:Gamma_HS} have been validated several times for dilute polymer solutions. In particular, Lee et al.~\cite{Lee2014,Lee2017} reported quantitative DO measurements in silica nanochannels induced by steady gradients of  neutral poly(ethylene glycol) (PEG), while Akdeniz et al.~\cite{Akdeniz2024} recently studied the DP migration of polystyrene (PS) particles and their advection  by DO in  unsteady PEG gradients  in a  microchannel.
Both experiments showed that eq~\eqref{eq:DP_general_conc} and eq~\eqref{eq:Gamma_HS} account for the observations with $R_i$ close to  polymer gyration radius $R_g$. Similar conclusions were drawn for dextran solutions using stopped-flow diffusion cells by Staffeld and Quinn~\cite{Staffeld1989} and  Nguyen et al.~\cite{Nguyen2022}, although with an exclusion length larger than $R_g$, possibly because of polydispersity and conformation of dextran in solution~\cite{Staffeld1989}. 

Observations with neutral molecular solutes are scarce, and often diverge from theoretical predictions.
Paustian et al.~\cite{Paustian:15} studied 
steady water/ethanol (EtOH) gradients imposed in microchannels using  hydrogel membranes. These experiments revealed DP migration of PS particles   with negligible advection by DO.
The observed migration, also referred to as  {\it  solvophoresis}, was directed 
toward low EtOH concentration, hinting at repulsive PS/EtOH interaction. They were performed in concentrated regimes, beyond the range of validity of eq~\eqref{eq:DP_general_conc},  and showed that the DP drift varies as $v_\mathrm{DP} \sim \nabla \log x$, $x$ being the molar fraction of EtOH.  
Lee et al.~\cite{Lee2017} later reported DO in silica nanofluidic channels induced by steady EtOH gradients. Measured DO flow rates were directed toward high EtOH concentration, also suggesting a repulsive EtOH/silica interaction. This is in contradiction with the known attractive interaction between EtOH and silica in the static configuration. Lee et al.~\cite{Lee2017} solved this paradox using molecular dynamic simulations showing that dynamical contributions at the nanoscale have to be included in eq~\eqref{eq:DP_general_conc} and eq~\eqref{eq:Gamma_HS}  to correctly account for the direction and magnitude of the slip flow. 
Finally, Williams et al.~\cite{Williams2020} studied the case of steady gradients of glucose imposed in  microchannels connecting reservoirs (similar to the Dunn chamber in Ref.~\cite{Gu2023}). 
Amine-modified PS particles did not experience DP  in the gradient, but were  advected by DO  driven by the interaction between glucose and the polymer-coated channel walls. The direction of the DO slip velocity $v_s$ demonstrated a priori attractive interactions, but $v_s$ did not vary linearly with the imposed gradient, suggesting that other mechanisms have to be included in the models such as  dependence of viscosity and surface heterogeneity~\cite{Williams2020}.

The main experimental difficulty for neutral solutes lies in the weakness of the expected DP/DO velocities in dilute solutions, which make them hard to measure and to distinguish from  other possible hydrodynamic flows.
Indeed, assuming molecular range of interactions $R_i =\SI{5}{\angstrom}$, the expected DO slip velocity is only $v_s \simeq \SI{30}{\nano\meter\per\second}$ for a concentration gradient $\nabla C = \SI{0.1}{\mol\per\liter\per\mm}$ in water according to eq~\eqref{eq:DP_general_conc} and eq~\eqref{eq:Gamma_HS}. 
This difficulty can be overcome by considering higher concentration gradients leading in that case
to DP/DO velocities that are of the same order of those observed in dilute electrolyte solutions. This however raises difficulties for modeling. On the one hand, variations of fluid properties with concentration should then be carefully accounted for. Notably, variations of density  $\rho(C)$ lead to buoyancy-driven flows that can affect measurements~\cite{Williams2020,Gu2018}. Moreover, variations of viscosity  $\eta(C)$ can transport Brownian particles
independently of DP/DO, through the so-called {\it viscophoresis}~\cite{Wiener2023,Khandan2025}.  
Finally, variations in the solute/solvent interdiffusion coefficient $D(C)$ affect the concentration gradients, and additional experiments are required to evaluate and relate it to DP/DO velocities. On the other hand, imposing sufficient gradients on experimentally manageable length scales requires to exit from the strict dilute limit, questioning the validity of eq~\eqref{eq:DP_general_conc}. Using a mechanical approach, Marbach {\it et al.}\ extended the theoretical models in the regime of high solute concentrations, and showed that the DO slip velocity $v_s$ is given by~\cite{Marbach2017}:
\begin{eqnarray}
    v_s =  - \Gamma \nabla \Pi,\label{eq:DPHS_Pi}
\end{eqnarray}
with a diffusio-osmotic mobility $\Gamma$ that depends both on solution viscosity $\eta$ and interaction potential $\mathcal{U}$ between the solute and surface, and thus possibly on concentration  $C$. In the specific case of a hard-sphere potential, the expression for $\Gamma$ derived by Marbach {\it et al.}\ reduces to eq~\eqref{eq:Gamma_HS}. The effect of high concentrations is then simply  accounted for by replacing the term $RT \nabla C$ by the gradient of osmotic pressure $\nabla \Pi$, as in standard transport models for which interactions play a role in addition to entropy alone.
Yet, such a relation still requires definitive experimental validation.

\begin{figure}[htbp]
\centering
\includegraphics{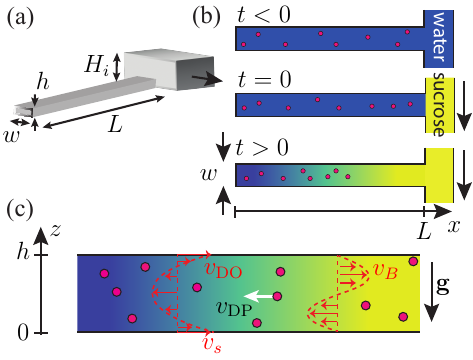}
\caption{DP migration of colloidal particles driven by a sucrose gradient in a dead-end microchannel. (a) Schematic 3D view of the dead-end pore geometry. (b) Sequence of experiments. The channels are initially filled
with water and dispersed colloids ($t < 0$).  At $t = 0$, a flow of sucrose
at a concentration of $C_0$ is imposed in the main channel. The sucrose
concentration gradient  in the channel for $t > 0$ leads to colloid
migration.
(c) Schematic section view showing the drift of particles by the superimposition of DP at velocity $v_\mathrm{DP}$, advection by DO at velocity $v_\mathrm{DO}$, and advection by the buoyancy-driven velocity field $v_B$. $v_s$ is the DO slip velocity at the channel walls and $\mathbf{g}$ indicates the gravity field.
\label{fig:Intro}}
\end{figure}

Among the various microfluidic geometries developed to date~\cite{Velegol2016,Marbach2019,Shim2022,Ault2025}, dead-end pore has emerged as a powerful and simple technique to study DP/DO, see, e.g., Refs~\cite{Kar2015,Shin2016,Akdeniz2023}.
The principle is sketched in Figure~\ref{fig:Intro}.
%%%%%%%%%%%%%%%%%%%%%%%
%%%%%%%%%%%%%%%%%%%%%%%
A dead-end channel with typical transverse dimensions $h \sim w \sim \SI{10}{\um}$ and length $L \sim \SI{1}{\mm}$  is connected to a wider  channel ($H_i \sim \SI{100}{\um}$) which in turn is connected to an inlet and an outlet. An appropriate sequence of flows imposed in the main channel allows us to induce temporal relaxation of a solute concentration gradient in the dead-end pore (Figure~\ref{fig:Intro}b). Beyond an entrance length on the order of $x_0  \sim h \ll L$, there is no net flow in the pore, and solute transport is dominated by diffusion. This is an advantage for studying DP/DO in a concentrated solution compared to other geometries with imposed flows, as it avoids the coupling between hydrodynamics and solute transport, which can be complex because of the variations of viscosity  $\eta(C)$ with local solute concentration.
 In turn, the colloids dispersed in the channel can be transported by DP and by DO flow induced at the channel walls  by the time-dependent concentration gradient, yet with no net flow in the channel (Figure~\ref{fig:Intro}c).  

In the present work, we used this geometry to study DP/DO induced by sucrose concentration gradients in relatively concentrated aqueous solutions, 
$C \simeq \SI{1}{\mol\per\liter}$, beyond the validity range of van't Hoff law $\Pi = RTC$. 
Despite   qualitative experimental evidence of DP induced by gradients of such solute~\cite{Paustian:15,VrhovecHartman2018},
there have been no quantitative studies on this question.
Furthermore, this issue of DP transport induced by sucrose gradients is relevant to many agro-industrial processes, in which high sucrose concentrations can be found, but also possibly in the context of sap transport in plants~\cite{Jensen2016}, sucrose being the main component of sap at concentrations of around $\SI{1}{\mole\per\liter}$.

We first designed  a dead-end pore geometry in a poly(dimethylsiloxane) (PDMS) chip in order to minimize parasitic flows, due to natural solutal convection and water permeation through the PDMS matrix. 
We then evidenced DP migration of colloidal PS particles up to velocities $v_\text{DP} \simeq \SI{3}{\micro\meter\per\second}$ directed against the sucrose concentration gradient ($\nabla C \simeq \SI{1}{\mol\per\liter\per\mm}$) suggesting repulsive interactions, and tracked their individual motion during the relaxation of the sucrose gradient. Using Raman confocal microspectrometry, we analyzed the time-dependent concentration field to estimate the interdiffusion coefficient $D(C)$ in the investigated range of concentration ($C = \SIrange{0}{1}{\mol\per\liter}$).   
Trajectories of individual particles can be modeled  using eq~\eqref{eq:DPHS_Pi}, with an exclusion length $R_i = \SI{5 \pm 0.9}{\angstrom}$ close to the characteristic size of  the sucrose molecule.
We finally showed that the spread of data cannot be accounted for solely by Brownian motion of the particles, suggesting  the possible advection of particles by a DO bulk flow in the channel.

\section{Materials and Methods}
\subsection{Physico-chemical properties of the aqueous sucrose solutions} 

%%%%%%%%%%%%%%%%%%%%%%%
\begin{figure}[htbp]
\centering
\includegraphics{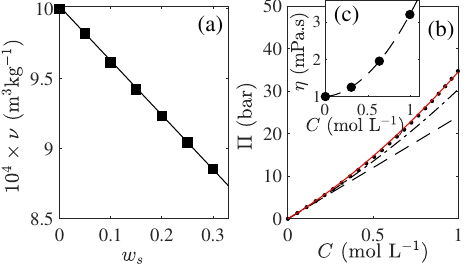}
\caption{Physical properties of aqueous sucrose solutions at $T=\SI{20}{\celsius}$. (a) Specific volume $\nu$ vs.\ mass fraction $w_s$~\cite{SucroseDataHandbook}. The continuous line is the fit by eq~\eqref{eq:volumespec}. (b) Osmotic pressure $\Pi$ vs.\ molar concentration $C$. Black dots:  literature data~\cite{Starzak1997}; dashed line:  van't Hoff law; dashed-dotted line: ideal solution; red line: fit by eq~\eqref{eq:Pivirial}. 
(c) Viscosity $\eta$ against $C$~\cite{Telis2007}. The dotted line is a second-order polynomial fit. \label{fig:DataSucrose}}
\end{figure}
%%%%%%%%%%%%%%%%%%%%%%%

All solutions were prepared with sucrose (Merck, molar mass $M_s = \SI{342.3}{\gram\per\mol}$) and deionized water from mass measurements up to sucrose mass fractions $w_s=0.3$. Figure~\ref{fig:DataSucrose}a shows the specific volume $\nu$ of the solutions  at $T=\SI{20}{\celsius}$~\cite{SucroseDataHandbook}. In the considered range of concentration, data are well described by:  
\begin{eqnarray}
    \nu = \nu_s w_s + \nu_w (1-w_s),
    \label{eq:volumespec}
\end{eqnarray}
with 
$\nu_w= \SI{1.002e-3}{\meter\cubed\per\kilo\gram}$ the 
specific volume of water and $\nu_s = \SI{6.135e-04}{\meter\cubed\per\kilo\gram}$.  This linear relation indicates the absence of volume change upon mixing in the studied concentration range.
In the following, we chose to present data against the sucrose molar concentration $C = w_s/(M_s \nu)$. The density of the solution $\rho = 1/\nu$ evolves linearly with $C$ according to:
\begin{eqnarray}
    \rho =\rho_w(1+\beta C), \label{eq:density}
\end{eqnarray}
in which $\rho_w = 1/\nu_w$ is the density of water, and $\beta = M_s (\nu_w - \nu_s) \simeq \SI{0.133}{\liter\per\mol}$ is the solutal expansion coefficient.

Figure~\ref{fig:DataSucrose}b shows the empirical correlation of osmotic pressure $\Pi$ of  aqueous sucrose solutions obtained by Starzak and Peacock from a comprehensive survey, rigorous in terms of thermodynamics and statistics, covering more than 50 data sets of water activity coefficients~\cite{Starzak1997}. 
In the concentration range investigated in the present work ($C=\SIrange{0}{1}{\mol\per\liter}$), data significantly deviate from the van't Hoff relation $\Pi = RTC$ and also from the osmotic pressure of an ideal solution $\Pi = RT\ln(1-x)/V_m$, $V_m$ being the molar volume of water and $x$ the sucrose mole fraction. In the following, we use the empirical relation:
\begin{eqnarray}
    \Pi = RT C (1+\epsilon \, C),\label{eq:Pivirial}
\end{eqnarray}
with $\epsilon \simeq \SI{0.4}{\liter\per\mol}$ to fit data from the literature. 

Figure~\ref{fig:DataSucrose}c displays data of viscosity $\eta$ of the solutions in the investigated range of sucrose concentration at $T=\SI{20}{\celsius}$. It shows an increase of the viscosity by a factor $\eta/\eta_w \simeq 3.2$ at $C=\SI{1}{\mol\per\liter}$,  $\eta_w$ being the viscosity of water.

\subsection{Microfluidic chip}

\begin{figure}[htbp]
\centering
\includegraphics{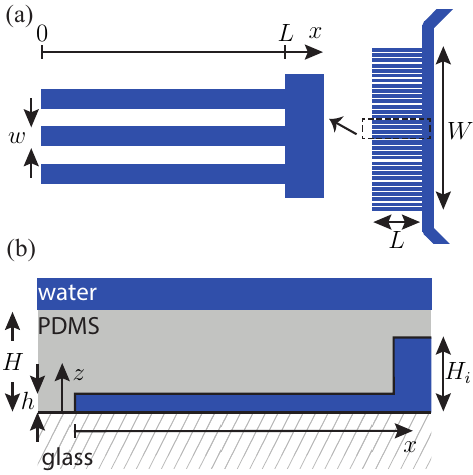}
\caption{Design of the microfluidic chip. 
(a) Top view:
$N=32$ dead-end channels of length $L=1$~mm and width $w=\SI{50}{\micro\meter}$ connected to 
a main channel. The  
distance between adjacent channels is   $\SI{50}{\um}$. The width of the inlet channel is $\SI{200}{\um}$.
(b) Side view evidencing the different heights: $h=\SI{55}{\um}$ for the Raman experiments shown in Figure~\ref{fig:RamanMap}, and
 $h=\SI{9}{\um}$ for the DP transport of particles. In both cases, $H_i=\SI{90}{\um}$, $H \simeq \SI{1}{\mm}$ and the chip is immersed in a water bath. \label{fig:FigSetup}}
\end{figure}

Figure~\ref{fig:FigSetup} shows the design of the microfluidic chip used in this study. It consists of an array of $N=32$ dead-end channels of length $L=\SI{1}{\mm}$ and width $w=\SI{50}{\um}$, all connected to a wider channel ($\SI{200}{\um})$ with one inlet and one outlet. The distance between adjacent channels is $\SI{50}{\um}$, and the overall width  of the  comb of dead-end channels is  $W=\SI{3.15}{\mm}$.   

The chips are made in poly(dimethylsiloxane) (PDMS) with standard soft lithography technique (Sylgard-184, mass ratio curing agent/polymeric
base $= 1/10$) from molds fabricated using a negative photoresist (MicroChem, SU-8). 
For the measurements of sucrose diffusion using Raman micro-spectroscopy, the height of the dead-end channels was $h=\SI{55}{\um}$ and the chip was sealed by plasma treatment with a thin glass slide ($\SI{170}{\um}$) to enable the use of an immersion objective lens.
For experiments on DP migration of particles in sucrose gradients, $h=\SI{9}{\um}$ and the chips were sealed on a thick glass slide ($\SI{1}{\mm}$). The channel height $h$ was measured using a 3D optical profiler (Neox, SensoFar) evidencing variations over the array of dead-end channels of the order of  $\simeq \SI{1}{\micro\meter}$  in the case $h=\SI{55}{\micro\meter}$, and $\simeq \SI{0.5}{\micro\meter}$ for $h=\SI{9}{\micro\meter}$.
In both cases, the height of the main channel was $H_i=\SI{90}{\um}$. 
%%%%%%%%%%%%%%%%%%%%%%%

%%%%%%%%%%%%%%%%%%%%%%%

\subsection{Minimizing pervaporation}
Water can permeate  through the matrix of a PDMS chip, a process known as {\it pervaporation}~\cite{Verneuil:04,Randall:05}, which leads to flows in the channels.
To estimate the order of magnitude of such flows that can interfere with the DP migration of colloids, we consider the case in which these flows  are a priori minimal, i.e., an infinitely thick PDMS chip, $H\to \infty$ in Figure~\ref{fig:FigSetup}. For simplicity, we assume Henry's law for water solubility in PDMS with a concentration at saturation $C_\mathrm{sat}$ ($\SI{}{\kg\per\meter\cubed}$), and a constant diffusion coefficient $D_w$  ($\SI{}{\meter\squared\per\second}$) of the water molecules in PDMS~\cite{Harley2012,Dollet2019}. The water pervaporation rate $Q_p$ 
($\SI{}{\meter\cubed\per\second}$)
from the array of channels of area $L\times W$  can be estimated by analogy with the diffusive heat transfer rate from an isothermal rectangular plate~\cite{Yovanovich2001}:  
$Q_p \simeq F\sqrt{L W} \tilde{q}  (1-\mathrm{RH})$ with $F \simeq 2.5$ a dimensionless shape factor, $\mathrm{RH}$ the ambient relative humidity, and $\tilde{q} = D_w C_\mathrm{sat}/\rho_w \simeq \SI{0.4}{\um\squared\per\second}$ ($C_\mathrm{sat} \simeq 0.7$~kg/m$^3$,
$D_w \simeq \SI{600}{\um\squared\per\second}$)~\cite{Harley2012,Dollet2019}.
For $\mathrm{RH} = 0.35$, $Q_p \simeq \SI{5}{\nano\liter\per\hour}$ leading to an entrance water flux  in each channel of the order of $v_p = Q_p/(N hw) \simeq \SI{100}{\nm\per\second}$ for $h=\SI{9}{\um}$. We confirmed this order of magnitude by measuring quantitatively the pervaporation rate at the scale of the channels, see Supporting Information, Video~S1. 

With no particular precaution, such flows would thus be close to the order of the DP drifts reported in this work.  In order to minimize them, the PDMS chips were immersed in a water bath during the experiments. This allows full stop of the pervaporation-induced flows in the channels, see Supporting Information, Video~S1. 
The transient time to saturate the PDMS matrix and thus stop pervaporation is of the order of $H^2/D_w \simeq \SI{30}{\min}$ for $H \simeq \SI{1}{\mm}$  according to reported values of $D_w$~\cite{Harley2012,Dollet2019}. Chips were therefore immersed during a few hours prior to the experiments to ensure that flow due to pervaporation was negligible.

\subsection{Microfluidic experiments  on diffusiophoretic migration}

We studied the migration of fluorescent PS particles (Thermo Fisher Scientific, yellow-green FluoSpheres) of two different diameters, $2a = \SI{500}{}$ and $\SI{1000}{\nm}$ with a polydispersity of $\pm \SI{15}{\nano\meter}$ according to the certificates of analysis provided by the manufacturer. Particles are dispersed in water at very low  volume fractions ($\varphi_0 \, \lesssim 10^{-6}$). We tested two different  surface functional groups,  sulfate- and  carboxylate-modified, and observed no significant differences in the results. 
All experiments were carried out at  room temperature, with $T=\SI{20}{\celsius}$.

Figure~\ref{fig:Intro}b schematically shows the course of the experiments carried out to study the  migration of the colloids in a sucrose gradient.
In the  first step, the dead-end channels are filled with the aqueous dispersion of fluorescent particles. Each channel (volume $h \, w \, L \simeq \SI{0.45}{\nano\liter}$, $h=\SI{9}{\um}$) contains water and approximately $10$ particles.
Next, we manually introduce an air bubble into the tube connected to the main channel, followed by a sucrose solution at concentration $C_0$ (in the following either $C_0 \simeq 0.99$ or $C_0 \simeq \SI{0.63}{\mol\per\liter}$ corresponding respectively to mass fraction $w_s=0.3$ and $0.2$).
The passage of the bubble  sets the time origin $t=0$ of the experiment, and results in an initial stiff gradient at $x \simeq L$.
A constant flow in the main channel is then imposed by a hydrostatic pressure drop of $\simeq \SI{20}{\milli\bar}$ resulting in a flow rate $Q\simeq \SI{200}{\micro\liter\per\hour}$ at $C_0 \simeq \SI{0.99}{\mol\per\liter}$ (estimated from  channel geometry  and viscosity of the sucrose solution, Figure~\ref{fig:DataSucrose}c) ensuring a constant concentration  at boundary $x=L$.

We finally acquired images with a 10X objective  (Evident, IX83) and an sCMOS camera (ORCA-Flash 4.0, Hamamatsu Photonics). The field of view of $\simeq 1.33\times 1.33 \SI{}{\mm\squared}$ allowed us to monitor simultaneously $13$ parallel channels.
The typical frame rate was $\SI{10}{frames\per\second}$ and  images were postprocessed using standard particle tracking algorithms~\cite{PTV}.

\subsection{Raman confocal spectroscopy \label{sec:RamanTech}}
Concentration fields of sucrose were measured along the dead-end channel using Raman confocal microspectroscopy. The setup consists of a Raman spectrometer (Andor Shamrock 303i, CCD camera iDus DU401-DD) coupled with an inverted microscope (Olympus, IX71) and a laser with a wavelength of \SI{532}{\nm}.

In the first step, we acquired Raman spectra of aqueous solutions of sucrose up to concentrations $C \simeq \SI{0.99}{\mol\per\liter}$ ($w_s=0.3$). 
These data were obtained using  a microscope objective  focusing the laser beam directly in glass vials containing the water/sucrose mixtures. 
The Raman scattered light collected by the same objective  is dispersed in the spectral range $\tilde{\nu}=200$-\SI{4000}{\per\cm}  with a grating of \SI{600}{lines\per\milli\meter} and a collection slit of width \SI{200}{\um}.  
%%%%%%%%%%%%%%%%%%%%%%%
\begin{figure}[htbp]
\centering
\includegraphics{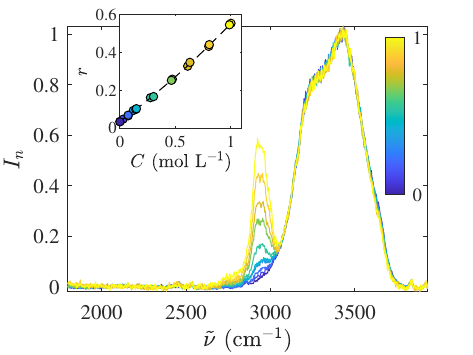}
\caption{Calibration of the Raman measurements. Normalized Raman spectra for different aqueous sucrose solutions in the concentration range $C=0$--$\SI{1}{\mol\per\liter}$ (blue to yellow). 
The inset shows the relative intensity $r$ of the Raman peak at $\tilde{\nu}=2920$--$\SI{2950}{\per\cm}$ due to sucrose. The dotted line is a second-order polynomial fit.
\label{fig:LibrarySpectra}}
\end{figure}
%%%%%%%%%%%%%%%%%%%%%%%
Figure~\ref{fig:LibrarySpectra} displays some of the measured Raman spectra  after correction of their baseline in the range $\tilde{\nu}=2000$-$\SI{2600}{\per\cm}$ and $\tilde{\nu}>\SI{3800}{\per\cm}$, and 
after normalization by the maximal intensity  of the Raman band of the OH stretching vibration at $\tilde{\nu} = \SI{3430}{\per\cm}$. A Raman peak specific to the vibration of sucrose emerges for $C>0$ in the range $\SIrange{2920}{2950}{\per\cm}$. The maximal intensity $r$ of this peak  increases from $r\simeq 0.036$ ($C=0$, pure water) up to $r \simeq 0.55$ at $C \simeq \SI{0.99}{\mol\per\liter}$, see the inset of Figure~\ref{fig:LibrarySpectra}. The standard deviation of these data, $\Delta r \simeq 0.003$, has been estimated by calculating  $r$ from spectra measured for fixed concentrations $C$ but in different experimental configurations (objectives 10X, 20X or 60X, acquisition time, measurement date), and is of the order of the size of the symbols in the  inset of Figure~\ref{fig:LibrarySpectra}.
Data $r$ vs.\ $C$ are well-fitted by a second-order polynomial, which can be used to determine the concentration $C$ of an unknown water/sucrose mixture from the analysis of its Raman spectrum. Experimental data are randomly dispersed around this calibration curve with a mean standard deviation  $\Delta 
C \simeq \SI{7}{\milli\mol\per\liter}$, which sets the typical standard uncertainty of these concentration measurements.

The measurements performed on-chip were obtained using a water immersion objective (magnification $60$X, NA$=1.2$) focusing the laser beam in the midplane of a single channel of height $h= \SI{55}{\um}$ with the same acquisition parameters as above, except for the use of a confocal pinhole (diameter $\SI{100}{\um}$) conjugated with the 
focal plane to minimize  out-of-focus contributions.  
The sequence of the experiment is the same as that  shown in Figure~\ref{fig:Intro}c but without fluorescent tracers dispersed in  water.
Raman spectra were continuously measured  along one of the dead-end channel by synchronizing the spectrometer with displacements of the chip using a motorized stage  (M\"arz\"auser Wetzlar). The spatial scan consisted of $11$ points from $x=\SI{50}{\um}$ to $x=\SI{1.05}{\mm}$, the last point being located in the main channel, and a complete scan took $\simeq \SI{21}{\second}$.

Despite  the confocal pinhole, spectra recorded in the channel display slight Raman peaks corresponding to the PDMS matrix, see Supporting Information, Figure~S1.
This contribution is superimposed with the sucrose peak at $\tilde{\nu}=\SIrange{2920}{2950}{\per\cm}$ and skew the estimation of $r$ for computing the sucrose concentration $C$ using the 
calibration curve of Figure~\ref{fig:LibrarySpectra}. 
However, we show in Supporting Information (Figure~S1) how this parasitic contribution can be subtracted from each spectrum and get a precise estimate of $r$.

 \section{Results}
\subsection{Particle tracking: evidence for migration \label{sec:expmigration}}

\begin{figure}[htbp]
\centering
\includegraphics{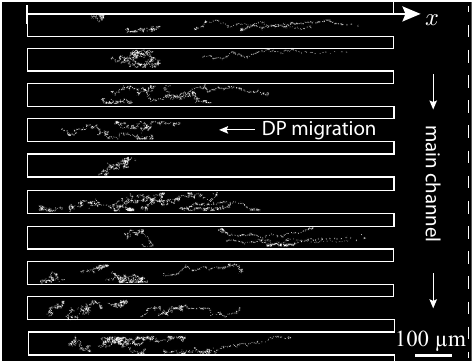}
\caption{Particles trajectories from superimposed images from $t=0$ to $t=\SI{465}{\second}$, see  Video~S2 in Supporting Information (particle radius
$a = \SI{250}{\nano\meter}$). Sucrose at a concentration
$C_0 = \SI{0.99}{\mol\per\liter}$ is imposed in the main channel at $t=0$.\label{fig:streakline}} 
\end{figure}

Figure~\ref{fig:streakline} shows particle trajectories in $10$ parallel channels for an experiment carried out 
at a sucrose concentration $C_0 \simeq \SI{0.99}{\mol\per\liter}$, superimposing images  
over a duration $\SI{8}{\min}$ after sucrose injection in the main channel at $t=0$ (Figure~\ref{fig:Intro}b), see also Supporting Information, Video~S2
(particle radius $a = \SI{250}{\nm}$).
%%%%%%%%%%%%%%%%%%%%%%%
%%%%%%%%%%%%%%%%%%%%%%%
Figure~\ref{fig:streakline} clearly evidence the migration of the colloidal particles toward the tip of the channels, i.e., down the sucrose concentration gradient. Particles initially located at the channel entrance ($X(t=0) \lesssim L$) initially drift at velocities  $\simeq \SI{3}{\mu\meter\per\second}$, which decrease over time until Brownian motion dominates for
$t \gtrsim \SI{20}{\min}$. Particles 
initially located at the channel tip ($x \gtrsim \SI{0}{}$) drift very little  and are subject only to thermal agitation.
Similar observations were reported for the two different particle sizes tested ($a = \SI{250}{}$ and $\SI{500}{\nm}$), and for imposed sucrose concentration $C_0\simeq \SI{0.63}{\mol\per\liter}$, but with smaller migration velocities in this case ($\simeq \SI{2}{\mu\meter\per\second}$). We also performed the reverse experiment corresponding to dead-end channels initially filled with sucrose and dispersed colloids and flushed the main channel with water at $t=0$, see Supporting Information, Video S3. In this case, particles are migrating toward the channel outlet, i.e., again down the sucrose concentration gradient.

These experiments suggest that the colloidal particles are subject to DP migration induced by the sucrose gradient with a repulsive interaction. To take these observations a step further, we now need to measure the sucrose concentration field to relate these observations quantitatively to the concentration gradient.

\subsection{Sucrose diffusion \label{sec:SucroseDiffusion}}

\begin{figure}[htbp]
\centering
\includegraphics{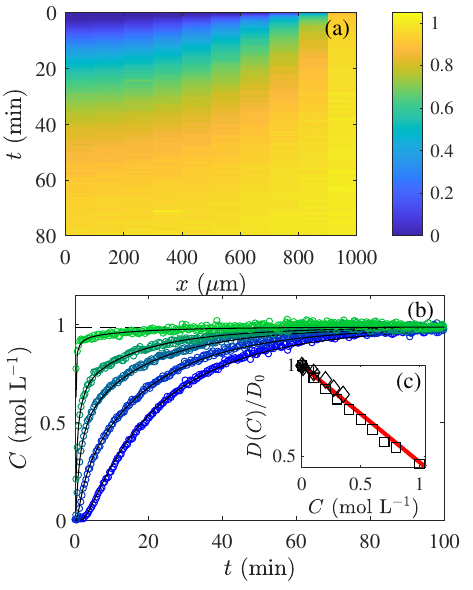}
\caption{Interdiffusion water/sucrose in the experiment shown in Figure~\ref{fig:Intro}c. (a) Space-time plot of the sucrose concentration field $C(x,t)$ coded in \SI{}{\mol\per\liter} with a colormap. (b) Temporal evolution of the concentration $C$ at $x=50$, $550$, $750$, $850$, and $\SI{950}{\um}$ (from blue to green). The continuous lines are the best fits by the diffusion models of eq~\eqref{eq:eqdiff} and eq~\eqref{eq:varD}. 
(c) Normalized diffusion coefficient $D(C)/D_0$ vs.\ $C$ at $T=\SI{25}{\celsius}$: $\star$: \cite{Ribeiro2006}, $\diamond$: \cite{Leaist1994}, $\square$: \cite{Irani1958}. The red line is eq~\eqref{eq:varD} that best fits the data shown in (b). 
\label{fig:RamanMap}}
\end{figure}

We performed  Raman confocal microspectroscopy to measure the sucrose 
concentration field in the experiment shown in Figure~\ref{fig:Intro}c. 
Due to the impact of the PDMS matrix, which interferes with the measured Raman spectra, these experiments were carried out with identical chips, but with a  channel height $h=\SI{55}{\micro\meter}$.
Figure~\ref{fig:RamanMap}a shows the space-time diagram of the sucrose concentration field $C(x,t)$  as a 2D map for an experiment
with an imposed concentration $C_0 \simeq \SI{0.99}{\mol\per\liter}$ in the main channel at $t=0$. Figure~\ref{fig:RamanMap}b represents the associated temporal relaxation of $C(x,t)$ at several positions $x$ in the channel.  These data show that significant concentration gradients develop within the channel until equilibration at times $t\geq \SI{40}{\min}$. A transport model is then needed to retrieve the sucrose/water interdiffusion coefficient from these data. 

%%%%%%%%%%%%%%%%%%%%%%%
%%%%%%%%%%%%%%%%%%%%%%%

Because the density of the aqueous solution of sucrose evolves with  concentration $C$, see eq~\eqref{eq:density}, concentration gradients in the channel are unavoidably associated with horizontal density gradients that in turn induce convection (Figure~\ref{fig:Intro}c).  The impact of this flow on the diffusion has been investigated in Ref.~\cite{Salmon2021} for microchannels of rectangular cross-section. It can be evaluated through the Rayleigh number:
\begin{eqnarray}
\mathrm{Ra} = \frac{\rho_w \beta C_0 g h^3}{\eta_w D_0}, \label{eq:RA}
\end{eqnarray}
with $g$ the acceleration due to gravity, $D_0 \simeq \SI{4.9e-10}{\meter\squared\per\second}$ the diffusion coefficient of sucrose at infinite dilution and $T=\SI{20}{\celsius}$~\cite{Mogi2007}.
This number compares the time scale  of convection over distance $h$ by the buoyant velocity $\sim \rho_w \beta C_0 g h^2/\eta_w$, with the time scale of diffusion over the channel height $\sim h^2/D_0$~\cite{Selva:12}.  
For $h=\SI{55}{\um}$ and $C_0 \simeq \SI{0.99}{\mol\per\liter}$, the Rayleigh number is $\mathrm{Ra} \simeq 440 < 10^3$, and  convection associated with density gradients remains negligible compared to diffusion whatever the extent of the density gradient according to Ref.~\cite{Salmon2021}. 

Because the specific volume of the water/sucrose solution evolves linearly with the sucrose mass fraction (Figure~\ref{fig:DataSucrose}a), there is no volume flow in the channel, and the concentration  is expected to follow a simple diffusion equation:
\begin{eqnarray}
    \frac{\partial C}{\partial t} = \frac{\partial}{\partial x} \left( D(C)\frac{\partial C}{\partial x} \right), \label{eq:eqdiff} 
\end{eqnarray}
with $D(C)$ the interdiffusion coefficient, possibly depending  on $C$. Such 1D description is fully valid because of the  high aspect ratio $L \gg w \sim h$ of the dead-end channels~\cite{Ault2018}. Eq~\eqref{eq:eqdiff} a priori applies all along the dead-end pore except at its entrance for $x \to L$, where flow in the main channel induces convection~\cite{Battat2019}. To take this into account, we apply to eq~\eqref{eq:eqdiff} the effective boundary condition $C(x=L_\mathrm{eff},t)=C_0$ at $x=L_\mathrm{eff}$ possibly slightly smaller than $L$. At $x=0$, there is a no-flux of solute boundary condition, $(\partial C / \partial x)_{x=0} =0$. 
Eq~\eqref{eq:eqdiff} along with these two boundary conditions was numerically solved for various $L_\mathrm{eff}$ and assuming an empirical relation:  
\begin{eqnarray}
&&D(C) = D_0(1-\delta  C), \label{eq:varD}
\end{eqnarray}
with $\delta$ a fitting parameter. 
 As displayed in Figure~\ref{fig:RamanMap}c, the results are satisfactorily described with $\delta = \SI{0.53}{\liter\per\mol}$ and $L_\mathrm{eff} = \SI{0.97}{\mm}$.  The exact value of $L_\mathrm{eff}$ has an influence on the fit only for values $x\to L$ close to the pore entrance.
This estimated value of $L_\mathrm{eff}$
is consistent with flow patterns at the entrance of the dead-end pores measured in Ref.~\cite{Battat2019} and with our observation of colloidal particles entering the dead-end channels from the main channel on a length of the order of
$x_0 \simeq \SI{30}{\micro\meter}$, see  Supporting Information, Figure~S2.

We did not find any values of diffusion coefficient $D(C)$ reported in the literature at $T=\SI{20}{\celsius}$, but 
 data of $D(C)/D_0$ at $T=\SI{25}{\celsius}$ from Refs.~\cite{Ribeiro2006,Leaist1994,Irani1958} (with $D_0 \simeq \SI{5.5e-10}{\meter\squared\per\second}$) almost collapse with our measurements performed at $T=\SI{20}{\celsius}$ (Figure~\ref{fig:RamanMap}c). In the following, we assume that the same values of $D(C)$ and $L_\mathrm{eff}$ apply to the experiments performed with channels of smaller height, $h= \SI{9}{\micro\meter}$ (Figure~\ref{fig:streakline}). This is justified as buoyancy-driven flows are again expected to be fully negligible ($\mathrm{Ra} \simeq 2$ for $C_0 \simeq \SI{0.99}{\mol\per\liter}$) and because flow patterns at the entrance of the dead-end channels did not display significant differences with the  $h= \SI{55}{\micro\meter}$ case. 
 Data  shown in Figure~\ref{fig:RamanMap} demonstrate the need for taking into account the variation of $D$ with  concentration $C$ to predict the sucrose concentration gradient in such experiments.

%%%%

\section{Analysis}

\subsection{DP drift}

\begin{figure}[htbp]
\centering
\includegraphics{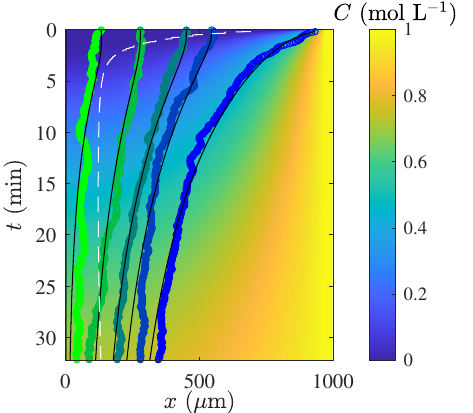}
\caption{\label{fig:Cxttraj} Sucrose concentration field computed from eq~\eqref{eq:eqdiff} and eq~\eqref{eq:varD} with $C_0  \simeq \SI{0.99}{\mol\per\liter}$ imposed in the main channel at $t=0$.  Five particle trajectories $X(t)$ with initial positions homogeneously distributed along the channel are also shown (from blue to light green). The continuous lines are the best fits of the trajectories by eq~\eqref{eq:Traj}. The dotted white line indicates $\mathrm{Pe}= 10$, see eq~\eqref{eq:Pec}.}
\end{figure}

Figure~\ref{fig:Cxttraj} shows different particle trajectories $X$ vs.\ $t$, tracked from images  such as the one displayed in Figure~\ref{fig:streakline}, superimposed to the predicted sucrose concentration field from eq~\eqref{eq:eqdiff} and eq~\eqref{eq:varD} (particle radius $a=\SI{250}{\nm}$, $C_0 \simeq \SI{0.99}{\mol\per\liter}$). 
%%%%%%%%%%%%%%%%%%%%%%%
%%%%%%%%%%%%%%%%%%%%%%%
Figure~\ref{fig:Cxttraj} evidences the clear correlation between the sucrose concentration gradient and the drift of the colloidal particles toward low sucrose concentration.
Multiple phenomena come into play to describe particle transport in the channel: DP, viscophoresis~\cite{Wiener2023,Khandan2025}, and advection by DO and buoyancy-driven flows.  However, we argue below that particle migration is a priori mainly driven by  DP due to the sucrose gradient. 

The interaction between the sucrose gradient and the
channel walls can induce a slip velocity $v_s$ at the solid surfaces, and thus a bulk flow in the channel (Figure~\ref{fig:Intro}c). This DO flow is also superimposed to the flow driven by the sucrose density gradient, even if the latter has no impact on the  sucrose concentration  field ($\mathrm{Ra} < 10^3$, see eq~\eqref{eq:RA}). Nevertheless, the cross-sectional average of these flows (volumetric net flux) is zero because the channel is dead-end~\cite{Salmon2021,Alessio2022}.
These flows cannot therefore explain the observed drift provided that Brownian motion ensures that particles explore the whole channel cross-section during their DP migration along $x$. This assumption is discussed later. To draw such a conclusion, we also implicitly assume that sedimentation and creaming do not play a role, i.e., that gravity does not affect the colloid distribution along $z$. This is justified even for 
the largest particle size ($a= \SI{500}{\nano\meter}$)
as the sedimentation/creaming length $\ell_s = k_B T/ (m^\star g)$ with $m^\star = \frac{4}{3}\pi [\rho_\mathrm{PS}-\rho(C)] a^3$ the buoyant mass of the  particles, varies from $\ell_s \simeq \SI{13}{}$ (pure water) to 
$-\SI{11}{\micro\meter}$ (sucrose concentration $C_0 \simeq \SI{0.99}{\mol\per\liter}$)  taking $\rho_\mathrm{PS} = \SI{1060}{\kilo\gram\per\meter\cubed}$ for PS density, particles being neutrally buoyant for $C\simeq \SI{0.45}{\mol\per\liter}$.
 Regarding viscophoresis (VP), the associated  drift can be estimated by $v_\mathrm{VP} = (\mathrm{d}D_c / \mathrm{d}x)$, with the colloid diffusivity given by the Stokes-Einstein relation:
\begin{eqnarray}
 D_c = \frac{k_B T}{6 \pi \eta(x,t) a} \label{eq:StokesEinstein}, 
\end{eqnarray}  $k_B$ being the Boltzmann constant and  $\eta(x,t)$ the local viscosity computed from the concentration field $C(x,t)$ shown in Figure~\ref{fig:Cxttraj}  and data of Figure~\ref{fig:DataSucrose}c. 
VP is expected to be negligible in our configuration as the maximal associated drift $ | v_\mathrm{VP} |$ remains below  $\SI{10}{\nano\meter\per\second}$ for $t > \SI{1}{\min}$ (see Supporting Information,  Figure~S4).

In the following, we thus assume that particles only drift because of DP and that $v_\mathrm{DP} = -v_s$ in eq~\eqref{eq:DPHS_Pi}, as particles are a priori significantly larger than the characteristic length of the interaction potential~\cite{Anderson1982}. Assuming negligible Brownian motion,  particle trajectories $X(t)$ thus verify: 
\begin{eqnarray}
    \frac{\text{d}X}{\text{d}t} = v_\mathrm{DP}(x,t)= \Gamma \frac{\partial\Pi}{\partial x}(X(t),t). \label{eq:Traj}
\end{eqnarray}
We solved numerically eq~\eqref{eq:Traj} with osmotic pressure $\Pi$ given by eq~\eqref{eq:Pivirial}, $C(x,t)$ computed from eq~\eqref{eq:eqdiff}, and $\Gamma$ as a fitting parameter to adjust the experimental trajectories such as those shown in Figure~\ref{fig:Cxttraj} using a least-squares minimization fitting procedure. 
We also considered that the uncertainty on particle location for each measured position $X(t)$ is dominated by Brownian motion and given by $\pm \sqrt{2 D_{c,0} t}$ with $D_{c,0}$ the colloid diffusivity estimated by eq~\eqref{eq:StokesEinstein} for $\eta = \eta_w$. This was used to weight each measurement point in the least-squares fitting, and eventually to estimate the relative standard uncertainty of the fit $\Delta \Gamma / \Gamma$ from the curvature at minimum of the chi-square estimation~\cite{NumericalRecipeC}, see Supporting Information for more details. Note that the choice $\eta = \eta_w$ provides an upper limit  of the  uncertainty on particle location. We checked that accounting for viscosity variations using eq~\eqref{eq:StokesEinstein} does not significantly change the results, because $D_c \sim 1/\sqrt{\eta}$ and the increase of $\eta$ is moderate.

The temporal integration of  eq~\eqref{eq:Traj} leads to satisfactory fits of the trajectories, with 
$\Gamma$ values for each trajectory and relative standard uncertainty $\Delta \Gamma / \Gamma$. We discarded trajectories with uncertainty $\Delta \Gamma / \Gamma > \SI{5}{\percent}$ which all correspond to particles with initial positions close to the channel tip at $x=0$ for which Brownian motion dominates over DP drift, see Supporting Information, Figure~S3. 
To rationalize this feature, we define the following local P\'eclet number that compares DP drift and colloid diffusion: 
\begin{eqnarray}
    \mathrm{Pe} = \frac{|v_\mathrm{DP}|x}{D_{c,0}}.
    \label{eq:Pec}
\end{eqnarray}
In eq~\eqref{eq:Pec}, $v_\mathrm{DP}$ is computed from eq~\eqref{eq:Traj} by taking the average value of $\Gamma$.
Figure~\ref{fig:Cxttraj} displays the isoline $\mathrm{Pe}= 10$ that delimits the transition from DP-dominated to Brownian-dominated transport. This transition line  roughly corresponds to the threshold of relative standard uncertainty of $\Delta \Gamma / \Gamma = \SI{5}{\percent}$  estimated numerically from the fits.

\begin{figure}[htbp]
\centering
\includegraphics{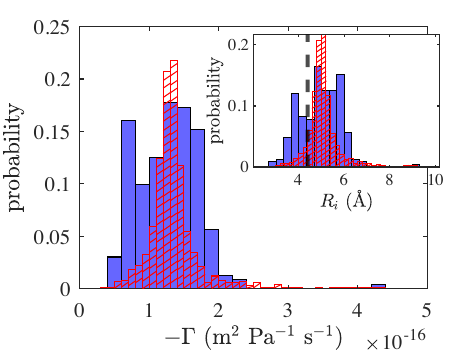}
\caption{\label{fig:Distrib_7manips} Probability distribution of diffusioosmotic mobility  $\Gamma$. Blue: experimental distribution  estimated from the fits of $230$ trajectories by eq~\eqref{eq:Traj}. Hatched red: distribution estimated from trajectories computed numerically to evaluate the role of Brownian motion, see eq~\eqref{eq:Langevin}.
Inset:  corresponding distributions of exclusion length $R_i$, see eq~\eqref{eq:Gamma_HS}. The vertical line is the Stokes radius of sucrose molecules,  $R_h  \simeq \SI{4.4}{\angstrom}$.}
\end{figure}

The distribution of $\Gamma$ values, omitting those from discarded trajectories, is centered around
$\overline{\Gamma} \simeq -\SI{1.3e-16}{\meter\squared\per\Pa\per\second}$ with relative standard deviation $\Delta \overline{\Gamma}/\overline{\Gamma} \simeq 0.35$  
(Figure~\ref{fig:Distrib_7manips}, blue).
It has been obtained from $\simeq 230$ selected trajectories, for two different particle radii ($a = \SI{250}{}$ and $\SI{500}{\nm}$), two different imposed sucrose concentrations ($C_0 \simeq 0.99$ and $\simeq \SI{0.63}{\mol\per\liter}$), and a reverse experiment (dead-end channels filled with sucrose, and water imposed in the main channel at $t=0$, see Supporting Information, Video~S3). 
No clear trend emerges from these experiments for the estimated $\Gamma$ values. 

In the case of a steric exclusion interaction between the sucrose and the surface of PS particles, modeled as a hard-sphere potential of range $R_i$, the diffusioosmotic mobility is given by eq~\eqref{eq:Gamma_HS}. 
 Note that viscosity variations of the solution 
(Figure~\ref{fig:DataSucrose}c) should not be considered into eq~\eqref{eq:Gamma_HS} for such potential, as the viscosity of the liquid layer with excluded sucrose molecules is  only water~\cite{Marbach2017}.
The inset of Figure~\ref{fig:Distrib_7manips} shows the distribution of $R_i$ estimated from the experiments and eq~\eqref{eq:Gamma_HS}. The average value $\overline{R_i} = \SI{5}{\angstrom}$ is close to the Stokes radius of sucrose $R_h  \simeq \SI{4.4}{\angstrom}$ estimated using $D_0 \simeq \SI{4.9e-10}{\meter\squared\per\second}$, and also  close to  the van der Waals radius $R_s=(3/(4\pi) \nu_s M_s / \mathcal{N}_a)^{1/3} \simeq \SI{4.4}{\angstrom}$ with  $\nu_s$ the specific volume in eq~\eqref{eq:volumespec} and  $\mathcal{N}_a$ the Avogrado number. The standard deviation is $\Delta R_i$  $\simeq \SI{0.9}{\angstrom}$  and includes both Stokes and van der Waals radii.
This consistency, together with the agreement of trajectories with fits along eq~\eqref{eq:Traj}, suggest that steric exclusion between sucrose and PS surfaces alone can explain the observed drift.
%%%%%%%%%%%%%%%%%%%%%%%
%%%%%%%%%%%%%%%%%%%%%%%

\subsection{Scatter of diffusioosmotic mobility}

Data shown in Figure~\ref{fig:Distrib_7manips} evidence non negligible scatter of the diffusiosmotic mobility $\Gamma$ and of the corresponding steric exclusion length $R_i$.
In the following, we explore possible sources for this spread only due to the particle transport: Brownian motion,  buoyancy-induced convection, and dispersion induced by DO.

\subsubsection{Brownian motion \label{ssec:Brownianmotion}}
Eq~\eqref{eq:Traj} describes the deterministic phoretic motion of colloids in a sucrose concentration gradient, but trajectories result from the superimposed random thermal agitation which could explain part of the observed scatter. Due to the space- and time-dependence of DP velocity field, estimating the spread related to thermal Brownian motion is not trivial. In order to explore this effect, we consider the 1D overdamped Langevin equation of the colloids motion:
\begin{equation}
    \frac{\dd X}{\dd t} =  \Gamma \frac{\partial \Pi}{\partial x}(X(t),t) + \frac{D_{c,0}}{k_B T}  F(t),
\label{eq:Langevin}
\end{equation}
with $\Gamma$ a constant diffusioosmotic mobility and $F$  a fluctuating force of zero average and with correlation:
\begin{equation}
    \langle F(t) F(t+\tau) \rangle) = 2 \frac{(k_B T)^2}{D_{c,0}}  \delta(\tau).
\end{equation}
\noindent with $\delta$ the Dirac distribution. For simplicity, we omit in eq~\eqref{eq:Langevin} the local variation of the diffusivity of the particles due to the change in viscosity with local concentration, and consider $D_{c,0}$  as the diffusion coefficient of the particles in water given by eq~\eqref{eq:StokesEinstein} for $\eta = \eta_w$.   

Eq~\eqref{eq:Langevin} has been solved numerically for $\Gamma = \overline{\Gamma}$  given by the mean value of the distribution shown in Figure~\ref{fig:Distrib_7manips}
and
$\simeq 2800 $ initial positions $X(t=0)$ ranging from $0.1L$ to $0.9L$,
and for the two particle sizes investigated ($a = \SI{250}{}$ and $\SI{500}{\nm}$). 
Then, we used the same weighted least-squares fitting procedure as for the experimental trajectories to extract a diffusioosmotic mobility $\Gamma$ from each individual simulated trajectory, and discarded data leading to $\Delta \Gamma / \Gamma > 5 \%$.
The corresponding distribution of fitted $\Gamma$ is plotted in  Figure~\ref{fig:Distrib_7manips} (red) along with experimental data.
Although Brownian motion along the $x$ axis significantly scatters the estimate of $\Gamma$ via trajectory fitting, the discrepancy between the two distributions shows that Brownian motion alone  cannot account for the observed experimental distribution.   

\subsubsection{Buoyancy- and DO-induced  dispersion \label{ssec:dispersionDObuoyancy}}

A possible DO bulk flow superimposed on the unavoidable convective flow driven by the density gradient along the channel can also transport the particles 
(Figure~\ref{fig:Intro}c). We emphasize that, as mentioned above, the associated volumetric net flux is zero because the channel is dead-end and these flows cannot explain the observed drift.
This is strictly true only when Brownian motion ensures that particles fully explore the channel cross-section during the characteristic time scale of their migration by DP given by $\sim D_0/L$ (Figure~\ref{fig:Cxttraj}).
This is almost verified in our experiments as $h^2/D_{c,0} \ll L/D_0$ and $w^2/D_{c,0} \sim L/D_0$, assuming uniform particle diffusivity $D_{c,0}$.
Advection of the particles by the associated shear flows can nevertheless interfere with their measured migration along $x$,  possibly leading to a broadening of the $\Gamma$ distribution obtained from trajectory fitting.  Such an effect is quantified through a P\'eclet number $\mathrm{Pe}_c = V h/D_{c,0}$, $V$ being a characteristic velocity scale of the considered flow. The influence of this advection can be discarded provided that this P\'eclet number is small enough, but the critical value should be evaluated carefully as it is not necessarily $1$.

The  impact of shear  on particle transport in microchannels can be evaluated quantitatively  using Taylor-Aris-like analyses~\cite{Young1991}.
This was  notably done by Alessio et al.\ to study the transport of colloids with superimposed DP and DO induced by salt gradients in a dead-end pore~\cite{Alessio2022} and by many other groups to study the role of buoyancy on solute transport in a wide range of experimental configurations, see, e.g., Refs.~\cite{Salmon2021,Erdogan1967,MACLEAN:01,Salmon2020}. In such analyses, the cross-sectional average  concentration field of particles is described in the lubrication approximation using a convection-diffusion equation with an effective diffusion coefficient $D_\mathrm{eff}$.
 This coefficient displays two contributions: one due to molecular diffusion $D_{c,0}$ and the other  which accounts for the longitudinal dispersion of the particles due to shear and that varies as $\sim \mathrm{Pe}_c^2$.

The application of such analyses to our experiments is not immediate due to viscosity variations, impacting both the particle diffusivity and the buoyancy-driven flows (Figure~\ref{fig:DataSucrose}c).
For simplicity, we will, nevertheless, assume that these variations play a minor role, and thus assume a uniform diffusivity $D_c \simeq D_{c,0}$ and viscosity $\eta \simeq \eta_w$.  
We also consider that the hypothetical DO slip velocity $v_s$ at the channel walls is not dependent on their nature (glass or PDMS). 
Under these assumptions, 
 the velocity fields associated with
DO and buoyancy have opposite parity, so  there is no coupling between these shear flows for the dispersion of a solute, see, e.g., Ref.~\cite{Salmon2020} for a derivation. 
In this case, the longitudinal dispersion of the particles along the channel by DO which scales as $\sim v_s^2 h^2/D_{c,0}^2$,
simply adds up to the one due to  buoyancy, which scales as
$\sim v_B^2 h^2/D_{c,0}^2$, with
$v_B$ the characteristic velocity scale of the buoyancy-driven flow~\cite{Erdogan1967}:
\begin{eqnarray}
    v_{B} = \frac{\rho_w   \beta g h^3 }{\eta_w} \frac{\partial C}{\partial x}. \label{eq:UB}
\end{eqnarray}
The detailed derivation of the exact effective diffusion coefficient can be found in many works, see, e.g., Refs.~\cite{Alessio2022,Erdogan1967,Salmon2020}, and the latter is written as:
\begin{eqnarray}
D_\mathrm{eff} =D_{c,0}\left(1+\frac{1}{\alpha}\frac{v_s^2 h^2}{D_{c,0}^2} + \frac{1}{\kappa}\frac{v_B^2 h^2}{D_{c,0}^2}\right), \label{eq:Deff}
\end{eqnarray}
($\alpha$,$\kappa$)  being numerical dimensionless factors that only depend on the aspect ratio of the channel cross-section. 
The above derivation implicitly assumes that secondary transverse flows due to buoyancy in the 3D channel~\cite{Erdogan1967} are negligible, an assumption fully valid in such microfluidic geometry~\cite{Salmon2021}.  
Eq~\eqref{eq:Deff} allows us to define quantitative criteria to evaluate the impact of buoyancy and the hypothetical DO flow on the dispersion of particles in the channel: $(v_B h/D_{c,0})^2 \ll \kappa$ and $(v_s h/D_{c,0})^2 \ll \alpha$. Note that numerical values of 
$\alpha$ and $\kappa$ are not trivial and given by $\alpha \simeq 210/6$  and $\kappa \simeq 4 \times 10^5$
for the considered aspect ratio\cite{Doshi1978,Chatwin1982,Salmon2021}, $h/w \simeq  9/50$. 

%%%%%%%%%%%%%%%%%%%%%%%
\begin{figure}[htbp]
\centering
\includegraphics{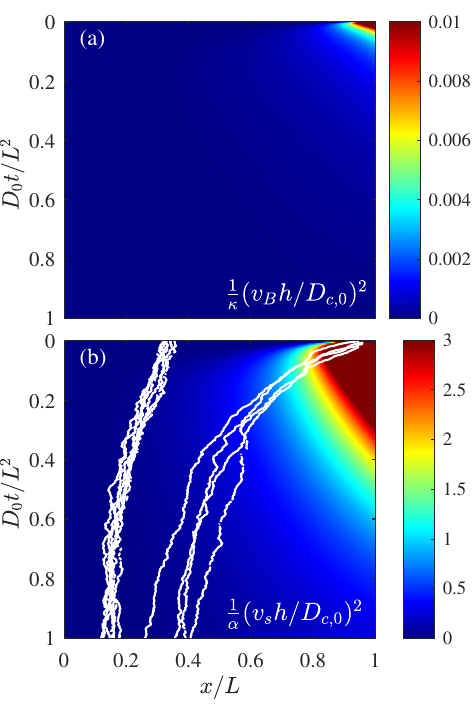}
\caption{Dispersion terms in eq~\eqref{eq:Deff}  due to buoyancy (a) and DO (b) coded with colormaps. In both cases, 
$a=\SI{250}{\nano\meter}$ and $C_0 \simeq \SI{0.99}{\mol\per\liter}$.  Experimental particle trajectories for initial positions $X(t=0)\simeq 0.35L$ and $\simeq 0.95L$ are also plotted in (b). 
\label{fig:dispersedtrajectories}}
\end{figure}
%%%%%%%%%%%%%%%%%%%%%%%

As far as buoyancy-driven dispersion is concerned, we computed the third term of  eq~\eqref{eq:Deff} for the particle size $a = \SI{250}{\nm}$, using eq~\eqref{eq:UB} and  solving eq~\eqref{eq:eqdiff} for the sucrose concentration field with $C_0 \simeq \SI{0.99}{\mol\per\liter}$. The numerical results shown in Figure~\ref{fig:dispersedtrajectories}a demonstrate that this term remains always small compared to $1$, and that  dispersion due to buoyancy can be  neglected. Note, however, that this term strongly varies with the channel height, $v_B^2 h^2/D_{c,0} \sim h^8$. Therefore, 
this conclusion  cannot be drawn if $h \geq \SI{20}{\micro\meter}$, even in regimes for which buoyancy does not impact the sucrose diffusion.

 Concerning the scattering of diffusioosmotic mobility $\Gamma$ due to dispersion in DO flow, as we have no observations of the possible slip  velocity $v_s$ at the channel walls, we will assume below that it is given by eq~\eqref{eq:DPHS_Pi} with the same average mobility coefficient $\overline{\Gamma}$ estimated from the DP drift. 
Figure~\ref{fig:dispersedtrajectories}b displays the DO term in eq~\eqref{eq:Deff}  computed
from the sucrose concentration field using eq~\eqref{eq:eqdiff} ($C_0 \simeq \SI{0.99}{\mol\per\liter}$,  particle size 
$a=\SI{250}{\nano\meter}$). This plot shows that the DO dispersion term is larger than $1$ when sucrose concentration gradients are high, i.e., at small time scales and close to the channel entrance. The same plot also shows experimental trajectories with initial conditions of $X(t=0) \simeq 0.35L$  and $\simeq 0.95L$  ($a=\SI{250}{\nano\meter}$ also for these trajectories).
Despite the  limited number of experimental trajectories it seems that the scatter of the trajectories of  particles that could be affected by DO at small time scales ($X(t=0) \simeq 0.95L$) is larger than for the particles with $X(t=0) \simeq 0.35L$  for which the dispersive term remains smaller than $1$.
This suggests the possible existence of a DO flow in the channel and its impact on the obtained distribution of diffusioosmotic mobility.

\section{Conclusions\label{sec:Conclusion}}

In this study, we have reported quantitative evidence of DP migration of colloidal particles driven by a time-dependent  concentration gradient of sucrose in relatively concentrated solutions, $C \simeq \SI{1}{\mol\per\liter}$.  Analyses of  individual trajectories and the sucrose concentration field allow us to demonstrate that a DP model  based on  steric exclusions with a hard-sphere potential over a length  $R_i = \SI{5 \pm 0.9}{\angstrom}$ explains the observed migration. Our data show the importance of taking into account both the variation of osmotic pressure $\Pi(C)$ and of the  interdiffusion coefficient $D(C)$ with concentration $C$, while minimizing parasitic contributions linked in particular to natural solutal convection. 

Analysis of the data suggests that a DO flow superimposed on DP migration could partly explain the experimental observed scattering of $R_i$.   
Note, however, that we did not observe any correlation between the measured instantaneous drift of the particles and their lateral $y$ position in the channel.
More precisely, we never observed significantly different drifts due to a hypothetical DO flow, when particles are close to the lateral channel walls at $y = \pm w/2$. This leads us to conclude that the magnitude of the possible DO slip flow $v_s$ at the channel walls is smaller than that of $v_\mathrm{DP}$. Also, the theoretical analysis leading in particular to eq~\eqref{eq:Deff}, is rigorously valid only for time scales larger than the particle diffusion time across the channel cross-section. While this condition is verified for the channel height in data presented in Figure~\ref{fig:dispersedtrajectories}, it is not strict for the channel width as one has $w^2/D_{c,0}$ close to $L/D_0$. A more accurate theoretical description of the dispersion in the regime $h^2/D_{c,0} \ll t < w^2/D_{c,0}$ are therefore needed to proceed further, see for instance Ref.~\cite{Ajdari:06}. Experiments in channels with even smaller cross-sections could also help clarify these points. However, the finite, non-negligible size of colloids in relation to cross-sectional dimensions could give rise to relevant issues relating to confinement and/or interactions between walls and particles. 

To conclude, there may also be other possible sources for the spread of $R_i$, e.g., physicochemical heterogeneities of the particles, and experiments enabling direct measurement of DO flow are needed to go further. Nevertheless, our results show unambiguously that a simple molecular interaction of the "exclusion" type can quantitatively account for particle migration in a sucrose gradient. Such observations could possibly be applied to macromolecules such as proteins  and be of importance in biophysics or industrial processes, given the ubiquity of sucrose in such cases. 

\section*{acknowledgement}
This research was partly funded by the French National Research Agency (ANR) as part of the ANR-23-CE30-0046 project. We also acknowledge Syensqo, CNRS, and Bordeaux University for financial support. 
BB is thankful for the support of the Iacocca International Internship program at Lehigh University. BB and HDO are supported in part by the US-NSF IRES Track I award 2153599.
JBS thanks F.~Doumenc and C.~Ybert for  insightful discussions. 
%\end{acknowledgement}

%%%%%%%%%%%%%%%%%%%%%%%%%%%%%%%%%%%%%%%%%%%%%%%%%%%%%%%%%%%%%%%%%%%%%
%% The same is true for Supporting Information, which should use the
%% suppinfo environment.
%%%%%%%%%%%%%%%%%%%%%%%%%%%%%%%%%%%%%%%%%%%%%%%%%%%%%%%%%%%%%%%%%%%%%
%\begin{suppinfo}

%Information about pervaporation (Video~S1), Raman spectra analysis, flow at the entrance to dead-end channels, error estimation procedure on mobility $\Gamma$, and viscophoresis.  
%Video~S2 corresponds to the experiment associated with Figure~\ref{fig:streakline}.
%Video S3 corresponds the reverse of the experiment shown in Figure~\ref{fig:streakline}.
%\end{suppinfo}

%%%%%%%%%%%%%%%%%%%%%%%%%%%%%%%%%%%%%%%%%%%%%%%%%%%%%%%%%%%%%%%%%%%%%
%% The appropriate \bibliography command should be placed here.
%% Notice that the class file automatically sets \bibliographystyle
%% and also names the section correctly.
%%%%%%%%%%%%%%%%%%%%%%%%%%%%%%%%%%%%%%%%%%%%%%%%%%%%%%%%%%%%%%%%%%%%%
%\bibliography{achemso-demo}
\bibliography{References}
\end{document}